# Learning of Content Knowledge and Development of Scientific Reasoning Ability: A Cross Culture Comparison

Lei Bao,[1, a] Kai Fang,[2] Tianfang Cai,[3]
Jing Wang,[1] Lijia Yang,[4] Lili Cui,[5] Jing Han,[1] Lin Ding,[1] and Ying Luo[6]

1. Department of Physics, The Ohio State university, OH, USA
2. Department of Physics, Tongji University, Shanghai, China
3. Department of Physics, Beijing Jiaotong University, Beijing, China
4. Department of Physics, National University of Defense Technology, Hunan, China
5. Department of Physics, University of Maryland at Baltimore County, MD, USA
6. Department of Physics, Beijing Normal University, Beijing, China

**Abstract:**

Student content knowledge and general reasoning abilities are two important areas in education practice and research. However, there hasn't been much work in physics education that clearly documents the possible interactions between content learning and the development of general reasoning abilities. In this paper, we report one study of a systematic research to investigate the possible interactions between students' learning of physics content knowledge and the development of general scientific reasoning abilities. Specifically, this study seeks to answer the research question of whether and to what extent content learning may affect the development of general reasoning abilities. College entrance testing data of freshman college students in both USA and China were collected using three standardized tests, FCI, BEMA, and Lawson's Classroom Test of Scientific Reasoning (Lawson Test). The results suggest that years of rigorous training of physics knowledge in middle and high schools have made significant impact on Chinese students' ability in solving physics problems, while such training doesn't seem to have direct effects on their general ability in scientific reasoning, which was measured to be at the same level as that of the students in USA. Details of the curriculum structures in the education systems of USA and China are also compared to provide a basis for interpreting the assessment data.

## I. INTRODUCTION

In most traditional education settings, teaching and learning emphasize predominantly training of content knowledge and it is often expected that consistent and rigorous content learning will help develop students' general reasoning abilities. However, there hasn't been much widely tested and accepted research evidence on the possible interactions between content learning in physics (or science and mathematics in general) and the development of general reasoning abilities. Most work in current physics education research also focuses on students' learning of content knowledge; although research on student general scientific abilities has been gaining popularity in recent years. There are productive new developments in instructional methods and materials that aim to foster students' scientific abilities.[1,2] In the experimental areas, several assessment studies have suggested a positive correlation between measures of students' abilities in scientific reasoning and mathematics and their achievement of content learning in physics.[3-5] On the theoretical side, a new theory on learning dynamics has been proposed which can provide a theoretical framework for understanding and modeling the interactions between general ability and content learning.[6,7] The theory also predicts a series of relations that can be tested by measurement. To advance our understanding in this area, it is important to conduct controlled experiments designed specifically to study the interactions between content learning and the development of general reasoning abilities.

The full scope of the possible interactions involves at least three basic research questions: (1) Whether and to what extent does content learning affect the development of general reasoning abilities? (2) Whether and to what extent does direct training on reasoning abilities affect content learning? (3) How will the interaction occur in and be affected by contexts? Here the contexts refer to a general definition that includes (but is not limited to) features of the content and the education settings. To answer these questions, systematic research has to be conducted in both theoretical and experimental areas. In this paper, we report one study that seeks to answer the first research question.

The research design makes use of the differences that naturally exist among different education systems. Students from different regions and education systems often go through very different training in science and mathematics courses during middle and high school years. This situation provides a natural setting of controlled variations on features of these trainings, which can be used to study the possible effects from such trainings on different aspects of student learning including content knowledge and reasoning abilities.

For example, in China, regular physics courses start from the 8$^{th}$ grade and continue through grade 12, providing five years of consistent training on most topics in the introductory physics area. The level of these courses is algebra based with serious emphasis on developing correct conceptual understandings and the skills to solve challenging problems without using calculus. As a result, the college level physics courses in China usually assume that students have developed understandings of most intro-level physics concepts and put more emphasis on calculus based physics problem-solving skills.

In contrast, school education in physics in the USA is more flexible. Students can have a choice to take one semester of elective physic course or they can also take more advanced AP physics courses. Obviously, the total time in training and the emphasis on conceptual development and problem-solving skills are all very different compared to that in China.

In addition to curriculum structure, the situations for science teachers are also very different. In the USA, well qualified science teachers are often in shortage in many states and there are comprehensive teacher training programs to help produce and maintain qualified science teachers. It is not uncommon for a high school science teacher to teach several different science courses. However, this versatility may sometimes limit the capability of a teacher to obtain a good mastery of teaching in a specific science discipline.

In China, several science topics are regularly taught in middle schools and high schools with varying emphasis. A non-exhaustive list includes physics (5 years), chemistry (3-4 years), geology (5-6 years), biology-health science (5-6 years), as well as a number of one- or two-semester short courses in certain technology and engineering areas such as computer and electronics, depending on local support. Teachers of these science courses are very specialized. Except in schools at some under-developed areas, it is not a common practice for a chemistry teacher to teach a middle school physics course or vise versa.

Most of the school science teachers in China graduate with bachelor degrees from the corresponding departments of Normal Universities. A Normal University is a comprehensive university with emphasis on producing quality teachers. Usually, more than half of the graduates from Normal Universities become K-12 school teachers. The level of undergraduate education on content knowledge in a discipline department of a Normal University is equivalent to what is required in the same department in any of the comprehensive universities in China. Therefore, it is safe to assume that most school teachers in urban areas are well equipped with the needed content knowledge and pedagogical skills. Thanks to the culturally based tradition of emphasis on education and the wide range of investments in the education infrastructure, China is able to produce about 10 million high school graduates annually. These students will go through the national college admission exam and approximately half of them will continue with some kind of higher education, although only a few percent can enter the top tier universities.

Because of the high competition, Chinese students go through a rigorous training in problem solving in all subject areas throughout most of their K-12 school years. It is well known that these students are good at solving very complicated content-based problems; however, the question of whether such training can bring the students into a high level of general skills and abilities that go beyond the specific content areas and problem types is a topic of common interests that hasn't been well studied.

In this paper, we report one study of a systematic research to investigate the possible impacts from differences in school training and cultural background on students' learning of content knowledge in physics and the development of more general abilities and skills in scientific reasoning. Specifically, we compare differences among freshman college students from USA and China on introductory mechanics, electricity and magnetism (E&M), and scientific reasoning abilities.

## II. RESEARCH DESIGN

The basic research issue in this investigation is to study whether and to what extent different training in middle and high schools on content knowledge in physics (or science and mathematics in general) may impact freshman college students' understandings of physics concepts and their ability in scientific reasoning. It is a big research question that requires many systematic detailed investigations. In this paper, we present a descriptive quantitative study aimed to document and contrast between USA and China the performances of freshman college students on standardized tests of physics concepts and scientific reasoning abilities. The results provide a baseline for further quantitative and qualitative studies and initial evidence on how training in middle and high school curricula many affect measures of students' content knowledge and reasoning ability.

The choice of contrasting US and Chinese students is not only because the Chinese student population is an interesting and less studied body but also because in China the middle and high school curricula on science (including physics) and mathematics is consistently controlled and enforced with a detailed national standard, which is also vastly different from that in USA. This natural setting provides a convenient research design, in which we define the training of physics knowledge in middle and high schools as an independent variable, and study the effects due to the variations of this variable in the natural settings of two very different education systems.

Although the primary variable is the training of physics content knowledge, the interpretation of the results can be extended to the training of content knowledge in science and mathematics in general. As discussed in the previous section, the content knowledge in science and mathematics courses in Chinese middle and high schools is taught at a substantially more sophisticated level than what is being taught in USA. If there exists any effect from the training of content knowledge to the development of scientific reasoning abilities, such effect should be considered as the results of the learning in science and mathematics in general.

The choice of comparing performances on physics tests is based on: (1) physics is one representative core piece of the science curricula, which is conceptually and logically sophisticated and is often considered to be related to students' ability of scientific thinking; (2) we have readily available research based standardized instruments to probe students' understandings of physics concepts.

It can be expected that if students' performances on standardized tests in math and other science topics can be compared, the Chinese freshman college students should also score better on these content oriented tests, since all these students have to score well in the related topics on the very challenging college entrance examination. An example of the level of math problems can be found in a recent BBC report. [8]

## A. Research Context

Data was collected from one US and two Chinese universities. All are major comprehensive universities of similar ranking in their respective countries. We will refer to the two Chinese universities as C1 and C2, and the US university as U1. All students tested are freshman science and engineering majors enrolled in calculus based intro-physics courses, and the tests were administered before any college level instruction on the relevant content topics.

Three standardized tests were used, which include FCI, [9] BEMA, [10] and the Lawson's Scientific Reasoning Test (Lawson Test).[11] The students in China used the Chinese versions, which were translated by physicists fluent in both languages. The translated versions were also piloted with a small group of undergraduate and graduate students (N~20) to remove apparent language issues.

Both FCI and BEMA are well known instruments in the physics education community. The Lawson's test is less widely used. It is designed to probe students' ability to apply aspects of scientific and mathematical reasoning, which include conservation of mass and volumes, proportional reasoning, control variables, probabilistic reasoning, correlational reasoning, and hypothetico-deductive reasoning.[11] A brief description of test items of the three instruments is included in the appendix.

## B. Data Collection

Test data was collected from over 3000 freshman students in the three universities. FCI data from U1 is an accumulated pool collected from 2005 to 2007. The rest of the data were all collected during 2007. Table 1 lists the population groups tested at the different universities and countries.

Table 1. Sample sizes of students taking different tests

| Universities | N | Tests |
|---|---|---|
| U1 | 1950 | FCI |
|    | 235 | BEMA |
|    | 646 | Lawson |
| C1 | 212 | FCI |
|    | 211 | BEMA |
|    | 206 | Lawson |
| C2 | 122 | FCI |
|    | 120 | BEMA |

## III. RESULTS AND ANALYSIS

Students' performances on these tests are summarized in Tables 2~4. A quick overview shows significant and large differences on FCI and BEMA results between Chinese and US students. On the other hand, students' performances on the Lawson Test are almost identical with the US students being slightly higher. The details of statistical analysis are given in table 5.

The data shows statistically significant variations among test results of all different populations on FCI and BEMA. By further comparing the effect size, it is easy to see that between US and Chinese students, there are categorical differences in student performances on tests of physics content knowledge. On the other hand, performances of the two populations on test of more general reasoning abilities are basically the same. This is an interesting result indicating that training on content knowledge might not have an obvious impact on students' development of general scientific reasoning abilities.

Table 2. FCI test results (30 items)

| Classes   | U1 (N=1950)  | C1 (N=212)   | C2 (N=122)   |
|-----------|--------------|--------------|--------------|
| Mean      | 14.8 (49.5%) | 25.5 (85.0%) | 26.6 (88.7%) |
| Median    | 14.0 (46.7%) | 26.0 (86.7%) | 27.0 (90.0%) |
| Std. Dev. | 5.8 (19.3%)  | 4.2 (14.0%)  | 3.3 (11.0%)  |

Table 3. BEMA test results (31 items)

| Classes   | U1 (N=235)  | C1 (N=211)   | C2 (N=120)   |
|-----------|-------------|--------------|--------------|
| Mean      | 9.7 (31.3%) | 19.9 (64.2%) | 21.1 (68.1%) |
| Median    | 9.0 (29.0%) | 21.0 (67.7%) | 22.0 (71.0%) |
| Std. Dev. | 2.9 (9.4%)  | 4.1 (12.6%)  | 3.8 (12.3%)  |

Table 4. Lawson test results (24 items)

| Classes   | U1 (N=646)   | C1 (N=206)   |
|-----------|--------------|--------------|
| Mean      | 18.3 (76.3%) | 17.9 (74.6%) |
| Median    | 19.0 (79.2%) | 19.0 (79.2%) |
| Std. Dev. | 4.2 (17.5%)  | 3.8 (15.8%)  |

To see more details of possible population differences, the score distribution of US and Chinese students on all three tests are plotted in Figures 1-3. In the plots for FCI and BEMA results, the data from the two Chinese universities are combined, since these two belong to the same larger population.

Table 5. Statistical analysis of student performances on FCI, BEMA and Lawson tests with US and Chinese first year college students. The p-value is computed based on a two-tailed t-test while the effect size is computed with the Hedges method.[12]

|  |  | C1 vs. U1 | C1 vs. C2 |
|---|---|---|---|
| FCI | p-value | 0.000 | 0.009 |
| FCI | Effect-size | 1.89 | 0.28 |
| BEMA | p-value | 0.000 | 0.008 |
| BEMA | Effect-size | 2.89 | 0.30 |
| Lawson | p-value | 0.201 |  |
| Lawson | Effect-size | 0.10 |  |

From the FCI results shown in Figure 1, we can see that the American students have a rather flat distribution in the middle score range (0.25~0.75) starting from the "chance" level, which is about ¼ of the total score. This is consistent with the school education system in the USA, which produces students with a diverse experience in physics learning: some had AP physics and had studied most of the mechanics concepts, some had regular elective physics courses and had studied some basic ideas and elements at different levels, and there are also some who didn't have any exposure to a formal physics course. The relatively flat distribution shows a somewhat uniform blend of students with different backgrounds in high school physics.

The Chinese students, on the other hand, all have gone through an almost identical nationally required physics curriculum spanning a 5-year period from the $8^{th}$ grade to the $12^{th}$ grade. This type of consistent training produces a narrow distribution that peaks at near the 90% score.

For the BEMA test (see Figure 2), the American students have a narrow distribution centered a bit above the chance level, indicating that most students didn't have much exposure to E&M concepts in their high school years. The Chinese students also scored lower than what they did on FCI and the score distribution is centered at around 70%. The results are again consistent with the effects of training during school years. The lower BEMA score from the Chinese students can be a result of the fact that E&M is usually a harder topic compared to mechanics and that some of the tested topics in BEMA (e.g., the Gauss' Law) are not taught or emphasized in the high school curriculum in China.

In the appendix, we have summarized the content areas taught in Chinese middle schools and high schools as well as the content topics measured in the FCI and BEMA. We can see that all the topics in the FCI are included in the Chinese middle- and high-school physics curricula, which are introduced at grade 8 and repeated at a higher level at grade 10. Most of the content topics tested in the BEMA are introduced at grade 9 and repeated at a higher level at grades 10 and 11.

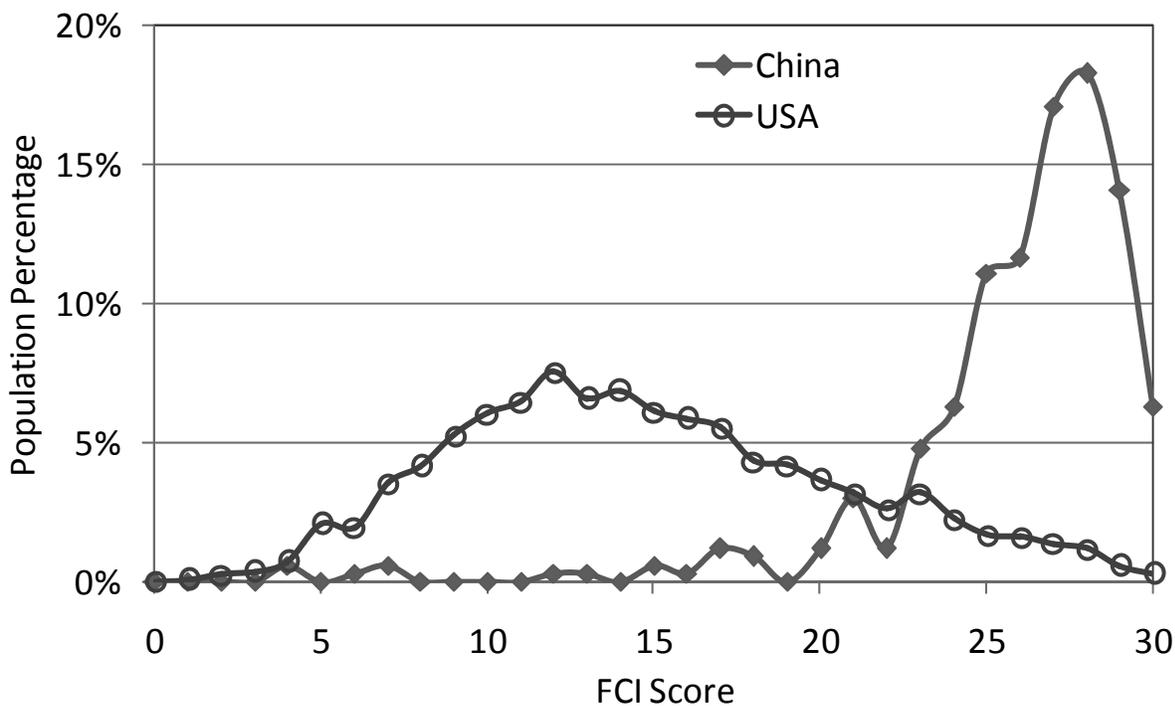

FIG 1. FCI college entrance score distribution of US and Chinese freshman college students. FCI is a 30-item test.

Summarizing the FCI and BEMA results from students in both China and USA, we can see that training in middle and high schools on the related physics content shows a direct impact on students' performance on these tests. A consistent rigorous training can put students into a fairly high performance level, while a flexible course structure with varying degree of emphasis often produce a diverse population distributed over a wide range of performance levels.

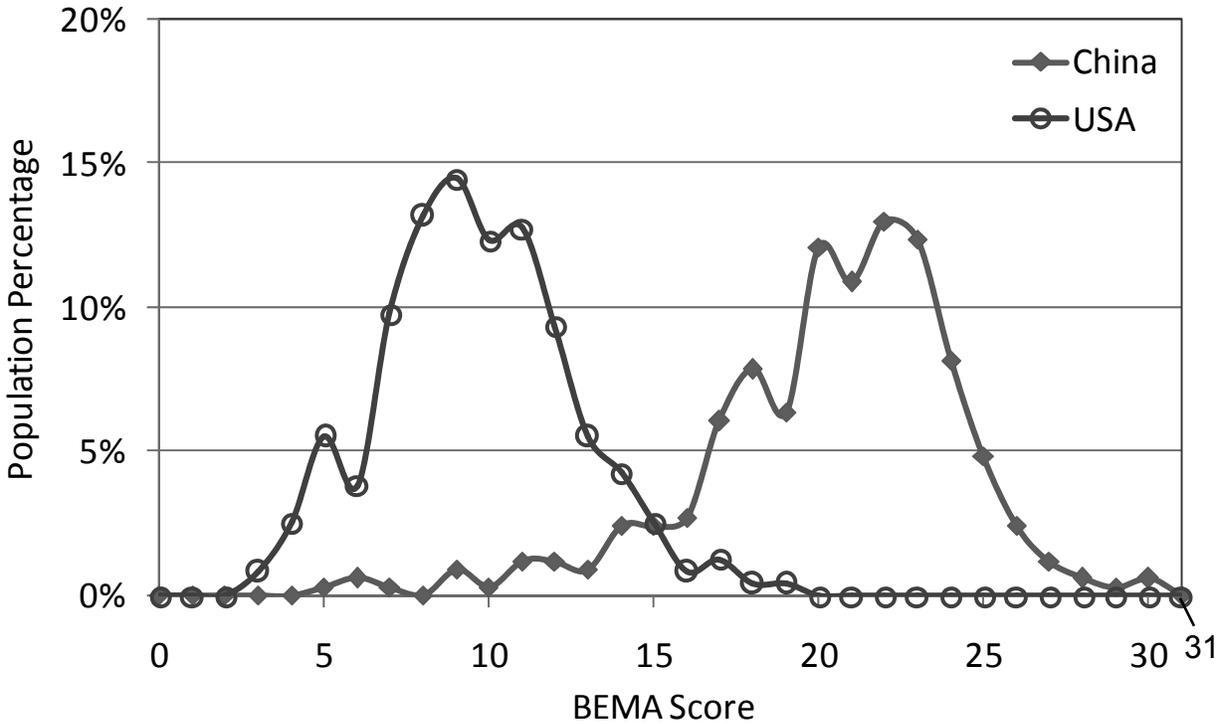

FIG 2. BEMA college entrance score distribution of US and Chinese freshman college students. BEMA is a 31-item test.

The results of the Lawson Test show a completely different pattern: the distributions of the Chinese and US students are almost identical. Compared to the content learning results measured with FCI and BEMA, it suggests that the large differences between the education systems in USA and China don't seem to cause much variation in students' scientific reasoning abilities measured with the Lawson Test.

To further explore the possible connection between training and students' development of content knowledge and reasoning abilities, we collected data with all three tests from students in one class in a Chinese University (C1). The correlations between students' scores on different tests are shown in table 6. We also included one US class' result for the correlation between students' test scores of Lawson Test and FCI.

Table 6. Correlations between test results on Lawson, FCI, and BEMA

| Classes | Lawson − FCI | Lawson − BEMA | FCI − BEMA |
|---|---|---|---|
| C1 (N=80) | 0.12 | 0.17 | 0.70 |
| U1 (N=111) | 0.20 | | |

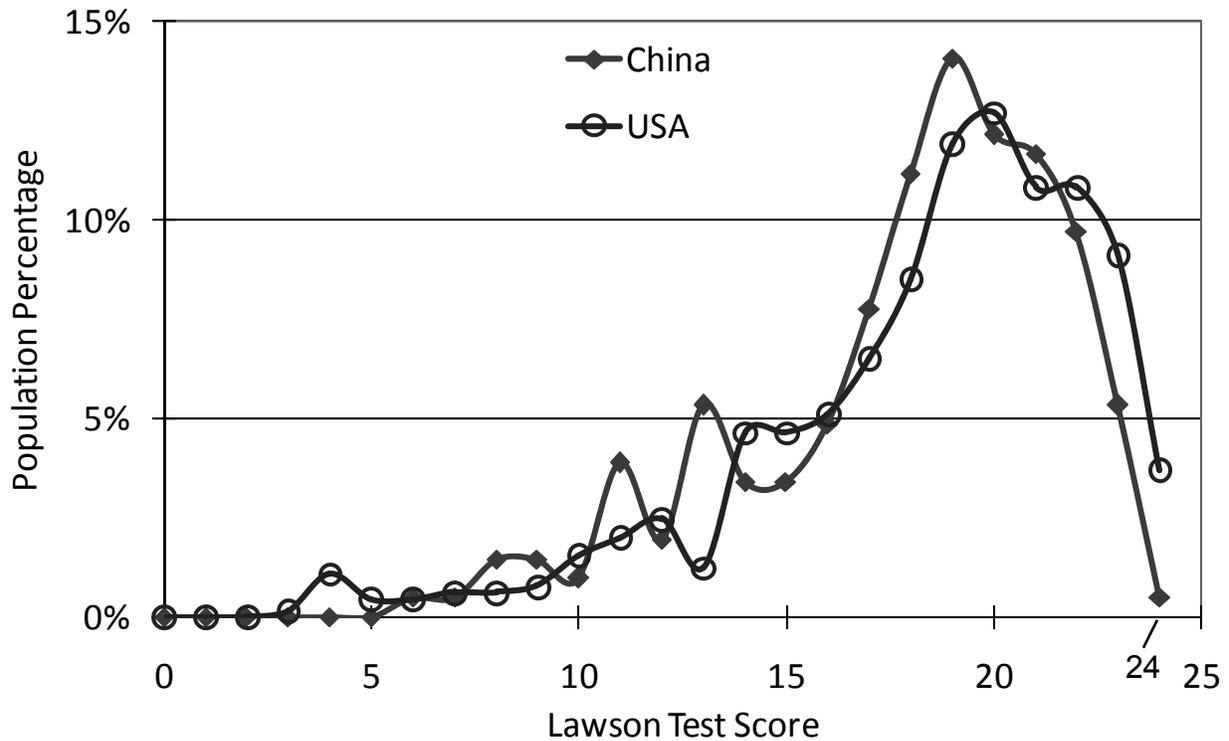

FIG 3. Lawson Test college entrance score distribution of US and Chinese freshman college students. Lawson Test is a 24-item test.

In general, the correlations between scores on Lawson Test and FCI or BEMA are small. On the other hand, a large correlation is observed between students' scores on FCI and BEMA. This result provides further evidence supporting that students' learning of content knowledge (under current education settings) measured with FCI or BEMA doesn't have an obvious connection to their development of general reasoning ability measured with Lawson Test. Meanwhile, similar training on content knowledge has produced highly correlated learning outcome on the two content areas of mechanics and E&M. These results are also consistent with the theoretical predictions about the correlation between measures of general ability and content knowledge. [6]

Comparing the two situations, we can see that under current education settings the learning of content knowledge and the development of general reasoning abilities do not have an obvious connection. Since only the content knowledge has been explicitly taught, this relation indicates the weak effect from content learning on the development of reasoning abilities. In this study, the research design doesn't address the question of whether and to what extent training of reasoning ability may impact content learning, which requires a completely different research design.

In summary, this study has produced evidence for answering the first research question; it is suggested that training in content knowledge doesn't seem to have an obvious effect on the development of general scientific reasoning abilities.

## IV. DISCUSSION

Since there are large differences in middle and high school physics courses between China and USA, and in both countries there are no explicit direct training of reasoning ability in school curricula, the naturally formed two education systems provide a unique controlled setting for studying the effects from content learning on the development of reasoning ability. The result suggests that training on content knowledge in physics (or science and mathematics in general) in current traditional school education settings, which is carried out at a substantial level of complexity in China, doesn't affect the development of general scientific reasoning abilities.

Although there are many additional uncontrolled variables such as social and cultural factors, the signal observed in this study by contrasting students' performances on both content and reasoning tests is large enough so that the uncontrolled variations wouldn't affect the conclusion. The fact that both the US and Chinese students have almost identical performances on the reasoning test indicates that on this ability measure (which are not directly taught in schools), the effects of the uncontrolled variables such as social and cultural factors don't seem to make a noticeable difference.

From a more general perspective, the performance differences between content knowledge (which is explicitly trained) and reasoning ability (which is not directly taught) indicate that in school education settings direct explicit teaching can have a powerful effect on students' learning, although the effectiveness of such training may vary with the format of education and population background. The results in this study have shown such effect on the learning of content knowledge. The next research question for future study is to find out whether direct training on reasoning abilities will make a similar impact.

One possible complication in this study is that the Lawson Test is a relatively simple test. The average score from typical college students is over 70%, suggesting possible ceiling effects. More detailed analysis on this instrument will be given in future papers. Here, we can make a simple analysis of the ceiling effects by comparing the distributions shown in Figure 3. We can see that when comparing lower half of the distribution, the two populations are still very similar. Since the scores of the lower half students are far from the ceiling, the similarity between these two sub populations of the two countries provides further evidence indicating that both populations are at about the same level regarding their performances on the Lawson Test.

It may come as a surprising result to see that training on physics content knowledge, which is presumably highly related to scientific logical reasoning, doesn't make any obvious impact on the development of students' reasoning abilities. A historically held belief among educators and researchers is that training in physics, which has a beautiful structure of logical and mathematical relations, would in general improve students' abilities in conducting reasoning that is intellectually challenging. However, the result from this study suggests that training in physics content knowledge in the traditional format alone is not enough to improve students' general reasoning abilities. The physics courses in Chinese middle and high schools emphasize predominantly traditional problem solving that is both conceptually challenging and computationally sophisticated. The research indicates that even content training at this level wouldn't transfer to affect the development of the general reasoning abilities.

Therefore, one has to seriously consider the education goals and methods, since the formats of teaching and what are being explicitly emphasized in the courses appear to have a dominant effect on what students can gain from these courses. This is in line with many current movements of education reforms that aim towards the constructivism paradigm.

Finally, we would like to summarize a few potentially useful carry-away messages for educators and researchers:

- Teaching and learning content knowledge in the traditional format often do not transfer to help students develop a good reasoning ability. Content knowledge and reasoning ability can be in two different categories that don't have to go together.

- New education methods are needed in order to help students develop both content understanding and reasoning ability. In order to do this, we need to look more seriously into the constructivism based education methods.

- To do better and more complete assessment of students' achievement, we need to use multiple types of measures including content knowledge, reasoning, and others (e.g., affective measures).

These messages are in support of the current education reforms and can help both researchers and educators moving forward on our effort in improving the current education systems.

**ACKNOWLEDGMENTS**

We wish to thank all the teachers helped with this research. Special thanks to Zuyuan Wang, Junjian Mao, and Yewen Zhang in the Department of Physics, Tongji University, Shanghai, China; Yingjian Qi, Communication University of China; Su Yang, Beijing Jiaotong University; Liming Li, Tsinghua University; Tianxing Cai, Beijing Technology and Business University. We also thank the anonymous reviewers whose careful comments have significantly improved the paper. This work is supported in part by the National Science Foundation under Grant Numbers DUE-0633473 and DUE-0618128. Any opinions, findings, and conclusions or recommendations expressed in this material are those of the author(s) and do not necessarily reflect the views of the National Science Foundation.

# APPENDIX

## A. Physics Education In Pre-University Section In China

In the Chinese middle and high schools, the students who want to choose science, engineering or other non-liberal arts areas as their majors in university are enrolling for about five years physics courses including of two years courses in the middle school and three years courses in high school. Tables A1~3 summarize the content areas included in these physics courses, compiled based on 2007 guideline published by the Ministry of Education of the People's Republic of China.

Table A1. Content themes of physics courses in Chinese middle schools and high schools [4]

| Sections | Themes |
| --- | --- |
| A | Force and Motion |
| B | Thermodynamics |
| C | Electricity and Magnetism |
| D | Waves |
| E | Optics |
| F | Nuclear physics |

Table A2. Section A: Force and Motion

| Categories | Concepts |
| --- | --- |
| Kinematics in one dimension | Reference frame, Displacement |
| | Velocity, average velocity, instantaneous velocity, speed |
| | Acceleration |
| | Motion with uniform acceleration |
| | Graphical analysis |
| | Falling bodies |
| | Acceleration due to gravity |
| Force | Concept of force and vector |
| | Force of gravity |
| | Elasticity and deformation |
| | Vector summation of forces |
| | Kinetic friction and static friction |
| Newton's laws of motion | Newton's first law of motion, inertia system |
| | Newton's second law of motion |
| | Newton's third law of motion |
| | Simple application of Newton's laws |
| | Units and SI system |
| Curvilinear motion | Vector of velocity in curvilinear motion |
| | Projectile motion |
| | General analysis of 2D motion |
| | Uniform circular motion |
| | Centripetal acceleration, centrifugal phenomena |
| | Newton's law of universal gravitation |
| Momentum | Momentum |
| | Law of conservation of momentum |
| | Center of mass |
| Work and Energy | Work |
| | Kinetic energy |
| | Potential energy |
| | Mechanical energy and conservation laws |
| Simple harmonic oscillations | Simple harmonic motion |
| | Amplitude, period, frequency of oscillation |
| | Energy in the simple harmonic oscillation |
| | Simple pendulum |
| | Forced vibration, resonance |

Table A3. Section C: Electricity and Magnetism

| Categories | Concepts |
| --- | --- |
| Electric field | Electric charges and charge conservation |
| | Coulomb's law |
| | Electric field, field lines |
| | Electric potential, electric potential energy |
| | Parallel-plate capacitor, capacitance |
| DC circuit | Ohm's law, voltage, resistance |
| | Phenomena of semiconductor and superconductor |
| Magnetism | Magnetic field and field lines |
| | Electric current and magnetic field |
| | Ampere's law |
| Electromagnetism | Phenomena of electromagnetism |
| | Faraday's law of induction |
| | Induction due to a moving conductor in magnetic field. |
| | Lenz's law |
| | Phenomenon of self-induction |
| AC circuit | Functions of resistance, capacitance and inductance in AC circuit |
| | Electric generator |
| | Theory of transformer |

## B. FCI Content Domains

Table A4. Content topics in FCI (summarized based on 1995 version)

|  | Inventory Item |
|---|---|
| 0. Kinematics | |
|     Velocity discriminated from position | |
|     Acceleration discriminated from velocity | 19E, 20D, (20E), (20A), (26A) |
|     Constant acceleration entails | |
|         parabolic orbit | |
|         changing speed | (3A) |
|     Vector addition of velocities | |
|     Circular motion | 5B,(5E), (6A),7B, (7E), 18B, |
| 1. First Law | |
|     with no force | 6B, 7B, 10A |
|         velocity direction constant | |
|         speed constant | |
|     with cancelling forces | 23B, 24A |
| 2. Second Law | |
|     Impulsive force | |
|     Constant force implies constant acceleration | 26E, |
| 3. Third Law | |
|     for impulsive forces | 4E, (4A), (4D) |
|     for continuous force | 15A, 16A, 25C, 28E |
| 4. Superposition Principle | |
|     Vector sum | 8B, 9E, 12B, 14D,(21B), 22B, 23B |
|     Cancelling forces | |
|     Force analysis | (11B), 17B, 29B, 30C |
| 5. Kinds or Force | |
|     5S. Solid contact | |
|         Passive | |
|         Impulsive | |
|         Friction opposes motion | 27C |
|     5F. Fluid constant | |
|         Air resistance | |
|         buoyant (air pressure) | (3E) |
|     5G. Gravitation | |
|         acceleration independent of weight | 1C, (1B), (2B) |
|         Free-fall | 3C, (3B), 13D, (13E) |
|         parabolic trajectory | 2A, (2D), (8D), 12B, 14D, 21E |

A parenthesis means the misunderstanding about the Newtonian concepts.

## C. BEMA Content Domains

Table A5. Content topics in BEMA.

| Categories | Concepts | Inventory Item |
|---|---|---|
| Electric field | Coulomb's law | 1, 2, 3 |
| | Electric field, field lines | 4, 5, 6 |
| | Electric potential | 14, 15, 16 |
| | Polarization of dielectrics | 7 |
| | Gauss' Law | 18 |
| | E-field in a conductor | 19 |
| Electric DC current | DC circuit | 10, 11, 17 |
| | RC circuit | 13 |
| | Model of resistivity | 12 |
| | Microscopic model of current | 8, 9 |
| Magnetism | Magnetic field and field line | 21, 22 |
| | B-field from current loops | 24 |
| | Charge in B and E cross fields | 26, 27 |
| | Magnetic force on a charge | 20, 23 |
| | Force between two parallel currents | 25 |
| | Force on moving charges – Hall effect | 30 |
| Electromagnetic Induction | Induction | 28, 29, 31 |

## D. Lawson's Test Content Domains

Table A6. Content topics in Lawson Classroom Test of Scientific Reasoning (ver. 2000)

| Question Pair | Question Type |
|---|---|
| 1, 2, 3, 4 | Conservation of Mass and Volume |
| 5, 6, 7, 8 | Proportional Thinking |
| 9, 10, 11, 12, 13, 14 | Control of Variables |
| 15, 16, 17, 18 | Probabilistic Thinking |
| 19, 20 | Correlational Thinking |
| 21, 22, 23, 24 | Hypothetico-deductive Reasoning |

**REFERENCES**

a) Email: lbao@mps.ohio-state.edu

[1] A. Boudreaux, P.S. Shaffer, P.R.L. Heron, and L.C. McDermott, "Student understanding of control of variables: Deciding whether or not a variable influences the behavior of a system," Am. J. Phys. 76 (2) 163-170 (2008).

[2] E. Etkina, A. Van Heuvelen, Suzanne White-Brahmia, David T. Brookes, Michael Gentile, Sahana Murthy, David Rosengrant, and Aaron Warren, "Scientific abilities and their assessment," Phys. Rev. STPER. 2, 020103 (2006).

[3] D. E. Meltzer, "The relationship between mathematics preparation and conceptual learning gains in physics: a possible "hidden variable" in diagnostic pretest scores," Am. J. Phys. 70(12), 1259-1268 (2002)

[4] V. P. Coletta & J.A. Phillips, "Interpreting FCI scores: Normalized gain, reinstruction scores, and scientific reasoning ability," Am. J. Phys., 73(12), 1172-1179 (2005).

[5] R.R. Hake, "Relationship of Individual Student Normalized Learning Gains in Mechanics with Gender, High-School Physics, and Pretest Scores on Mathematics and Spatial Visualization," Physics Education Research Conference; Boise, Idaho; August, 2002; online at <http://www.physics.indiana.edu/~hake/PERC2002h-Hake.pdf>.

[6] L. Bao, "Dynamic Models of Learning and Education Measurement," submitted to Phys. Rev. STPER, (2005); online at <http://arxiv.org/abs/0710.1375>.

[7] D. E. Pritchard, Y. Lee, and L. Bao, "Mathematical Learning Models that Depend on Prior Knowledge and Instructional Strategies," Phys. Rev. STPER, (in press)

[8] BBC News report, "Mathematicians set Chinese test," http://news.bbc.co.uk/2/hi/uk_news/education/6589301.stm

[9] D. Hestenes, M. Wells and G. Swackhamer, "Force concept inventory," The Physics Teacher, Vol.30, March, 141-158 (1992). Test used is the 1995 version.

[10] L. Ding, R.Chabay, B. Sherwood and R. Beichner, "Evaluating an electricity and magnetism assessment tool: Brief electricity and magnetism assessment," Phys. Rev. STPHYS. 2, 010105 (2006).

[11] A. E. Lawson, The development and validation of a classroom test of formal reasoning, J. Res. Sci. Teach. 15(1), 11-24, (1978). Test used in study: Classroom Test of Scientific Reasoning, revised ed. (2000).

[12] L. V. Hedges, and I. Olkin, *Statistical methods for meta-analysis*. San Diego, CA: Academic Press (1985).